\documentclass[preprint,aps,prb,showpacs,superscriptaddress]{revtex4}
\usepackage{graphicx}
\usepackage{amssymb}
\usepackage{bm}
\begin{document}
\title{Study of axial strain induced torsion of single wall carbon nanotubes by $2$D continuum anharmonic anisotropic elastic model}
\author{Weihua MU}
\affiliation{Institute of Theoretical Physics, The Chinese Academy
of Sciences, P.O.Box 2735 Beijing 100190, China}

\author{Ming Li}
\affiliation{Graduate University of Chinese Academy of Sciences,
 Beijing 100190, China}

\author{Wei Wang}
\affiliation{College of Nanoscale Science and Engineering (CNSE),
\\University at Albany, State University of New York, NY 12203, USA}
\author{Zhong-can Ou-Yang}
\affiliation{Institute of Theoretical Physics, The Chinese Academy
of Sciences, P.O.Box 2735 Beijing 100190, China} \affiliation{Center
for Advanced Study, Tsinghua University, Beijing 100084, China}
\begin{abstract}
Recent molecular dynamic simulations have found chiral
single wall carbon nanotubes (SWNTs) twist during
stretching, which is similar to the motion of a screw.
Obviously this phenomenon, as a type of
curvature-chirality effect, can not be explained by usual
isotropic elastic theory of SWNT. More interestingly, with
larger axial strains (before buckling), the axial strain
induced torsion (a-SIT) shows asymmetric behaviors for
axial tensile and compressing strains, which suggests
anharmonic elasticity of SWNTs plays an important role in
real a-SIT responses. In order to study the a-SIT of
chiral SWNTs with actual sizes, and avoid possible
deviations of computer simulation results due to the
finite-size effect, we propose a $2$D analytical continuum
model which can be used to describe the the SWNTs of
arbitrary chiralities, curvatures, and lengthes, with the
concerning of anisotropic and anharmonic elasticity of
SWNTs. This elastic energy of present model comes from the
continuum limit of lattice energy based on Second
Generation Reactive Empirical Bond Order potential
(REBO-II), a well-established empirical potential for
solid carbons. Our model has no adjustable parameters,
except for those presented in REBO-II, and all the
coefficients in the model can be calculated analytically.
Using our method, we obtain a-SIT responses of chiral
SWNTs with arbitrary radius, chiralities and lengthes. Our
results are in reasonable agreement with recent molecular
dynamic simulations. [Liang {\it et. al}, Phys. Rev. Lett,
${\bf 96}$, 165501 (2006).] Our approach can also be used
to calculate other curvature-chirality dependent
anharmonic mechanic responses of SWNTs.
\end{abstract}
 \maketitle
\maketitle 
The amazing mechanical properties of carbon nanotubes
(CNTs), such as high elastic modulus, exceptional
directional stiffness, and low density, make them idea
candidates for the applications of nanoelectromechanical
systems (NEMS)
devises~\cite{baughman02,craighead00,sapmaz03}. Recent
studies have demonstrated the possibilities of using CNT
as actuator~\cite{baughman99}, nanotweezers~\cite{kim99},
and nanorelay~\cite{kinaret03,lee04,jang05,jang08}.
Detailed understanding of mechanical behavior, especially
structurally-specific mechanical properties of CNT-based
NEMS devises is therefore crucial for their potential
applications in NEMS.

Unlike isotropic elastic thin shell, due to special geometries of
SWNTs, e.g. chiralities, there is coupling between axial strain and
torsion strain, which is similar to ordinary helical
spring~\cite{gartstein03}. More interestingly, recent molecular
dynamic simulations found asymmetric behaviors of such coupling in
chiral SWNTs~\cite{liang06,geng06} and double walled carbon
nanotubes (DWNTs)~\cite{zhang08}, namely, asymmetry of a-SIT for
tensile and compression strains~\cite{liang06}. Later, Upmanyu {\it
et al.}'s finite element method simulation also obtained asymmetric
a-SIT response~\cite{upmanyu08}. Main property of asymmetric a-SIT
is that a-SIT responses for tension or compression are much
different at large strain. Torsion angle per unit length increases
when strain increases in tension case. However, with increasing
strain under compression, the torsion angle firstly increases, then
decreases to zero, and increases again after changing the direction
of twist~\cite{liang06,geng06,zhang08}.

A-SIT implies the coupling between axial vibration modes and
torsional ones for chiral SWNTs, which may play an important role in
applications of CNT-NEMS oscillators~\cite{sazonova04,zhao03}. 
To understand a-SIT response, there are very few studies:
Gartstein, {\it et al.}~\cite{gartstein03} used a two-dimensional
continuum elastic model, predicted linear a-SIT effect for chiral
SWNTs with small strain, i.e., SWNT twists in opposite directions
for tension and compression and rotation angle varies linearly with
strain. Gartstein {\it et al} found a-SIT response is chirality
dependent, it reaches the maximum when the chiral angle is $\pi/12$.
Liang {\it et al.}'s molecular simulations~\cite{liang06} extended the
study of a-SIT to large
strain region (before buckling), obtained asymmetric a-SIT.
By comparing a-SIT response with changes of geometry of carbon-carbon bonds,
they found asymmetry a-SIT is relevant to microscopic lattice structure of SWNT. Geng
{\it et al}~\cite{geng06} studied both of torsion induced by axial
strain and axial strain induced by torsion, and showed nonlinear
axial stress-strain relation occurring in the same time. Upmanyu
{\it et al}'s~\cite{upmanyu08} finite element method simulation also obtained a-SIT.

All these efforts are valuable in understanding a-SIT.
Nevertheless, Gartstein {\it et al}'s theory was restrict
to linear a-SIT response, while the molecular dynamic or
finite element method simulations for a series of SWNTs
with some special chiral index were time-consuming, a lot
of computer resource were needed, which limits their
further application to study of properties of actual
SWNTs. Also, there is a general question for these
simulation work: can the results of the simulations for
small systems be extrapolated to SWNTs at equilibrium
state with actual sizes?

In our knowledge, there lacks a easily handled theoretical
frame capturing basic physics of asymmetric a-SIT which
can obtain this response for actual SWNTs at equilibrium
states with arbitrary radius and chiralities. To fulfill
this task, we propose a quasi-analytical approach based on
continuum elastic theory. In our model, the carbon-carbon
interactions in SWNT are described by REBO-II
potential~\cite{REBO02}, which is a classic many-body
potential for solid carbon and hydrocarbons. The
advantages for REBO-II potential are it has analytical
form of carbon-carbon pair potentials with the bond length
and bond angle as variables of energy functions, the
parameters of REBO-II potential were fitted from a large
data sets of experiments and {\it ab initio} calculations.
REBO-II potential can accurately reproduce elastic
properties of diamond and graphite, In
Ref.~\onlinecite{liang06}, molecular dynamic simulation
was also based on REBO-II potential.

The carbon-carbon interaction energy near the equilibrium state
without deformations can be obtained analytically by Taylor
expansion with inclusion of the most important cubic term, i.e., anharmonic term of bond stretching,
\begin{eqnarray}
\label{rebo}
V & = & V_{0}+\frac{1}{2}\sum_{\langle ij \rangle )}\left(\frac{\partial^{2}V}
{\partial r_{ij}^{2}}\right)_{0}\left(r_{ij}-r_{ij}^{0}\right)^{2}\nonumber\\
  &  & +\sum_{\langle ij \rangle}\sum_{k\neq i,j}\left(\frac{\partial^{2}V}{\partial r_{ij}\partial\cos\theta_{ijk}}\right)_{0}\left(r_{ij}-r_{ij}^{0}\right)\left(\cos\theta_{ijk}-\left(\cos\theta_{ijk}\right)^{0}\right) \\
 &  & +\frac{1}{2}\sum_{\langle ij \rangle}\sum_{k\neq i,j}\left(\frac{\partial^{2}V}{\partial\left(\cos\theta_{ijk}\right)^{2}}\right)_{0}\left(\cos\theta_{ijk}-\left(\cos\theta_{ijk}\right)^{0}\right)^{2}\nonumber \\
 &  & +\sum_{\langle ij \rangle}\sum_{k,l\neq i,j}\left(\frac{\partial^{2}V}
 {\partial\cos\theta_{ijk}\partial\cos\theta_{ijl}}\right)_{0}\nonumber\\
 & &\cdot\left(\cos\theta_{ijk}-\left(\cos\theta_{ijk}\right)^{0}\right)\left(\cos\theta_{ijl}-\left(\cos\theta_{ijl}\right)^{0}\right)\nonumber\\
 &  & +\frac{1}{3!}\sum_{\langle ij \rangle}\left(\frac{\partial^{3}V}{\partial r_{ij}^{3}}\right)_{0}\left(r_{ij}-r_{ij}^{0}\right)^{3}.\nonumber \end{eqnarray}
Here $\langle ij \rangle$ denotes the nearest neighboring
atom pairs, $\theta_{ijk}$ denotes angle between bonds
$i-j$ and $i-k$. Equilibrium state is denoted by $"0"$.
Similar series expansion of quadratic terms for Brenner
potential~\cite{brenner90} have been reported by Huang
{\it et al.}~\cite{huang06}.

The non-crossing second and fourth terms in right hand of Eq.
\ref{rebo} were also presented in Lenosky's model~\cite{lenosky92}.
From analytical form of REBO-II potential, the derivatives are,
\[
\left(\frac{\partial^{2}V}{\partial r_{ij}^{2}}\right)_{0} \approx
43.67\mathrm{eV\cdot\AA^{-2}},\;\left(\frac{\partial^{2}V}{\partial
r_{ij}\partial\cos\theta_{ijk}}\right)_{0}\approx-5.924\mathrm{eV\cdot\AA^{-1}},
\]

\[
\left(\frac{\partial^{2}V}{\partial\left(\cos\theta_{ijk}\right)^{2}}
\right)_{0}\approx3.187\mathrm{eV},\;\left(\frac{\partial^{2}V}{\partial\cos\theta_{ijk}
\partial\cos\theta_{ijl}}\right)_{0}\approx-0.367\mathrm{eV},\]
and

\[\left(\frac{\partial^{3}V}{\partial
r_{ij}^{3}}\right)_{0}\approx-333.4\mathrm{eV\cdot\AA^{-3}.}\]

In $2$D elastic theory of SWNT, the in-plane deformations of SWNT
can be described by~\cite{tu02}
\[
\underline{\bm{\varepsilon}}=\left(\begin{array}{cc}
\varepsilon_{1} & \varepsilon_{6}/2\\
\varepsilon_{6}/2 & \varepsilon_{2}\end{array}\right),\]
with
$\varepsilon_{1}\equiv\varepsilon_{11},\;\varepsilon_{2}\equiv\varepsilon_{22},
\;\varepsilon_{6}\equiv2\varepsilon_{12},$ are the axial,
circumferential, and shear strains, respectively. After
deformation, the bond vector from atom $i$ to its three
nearest neighboring atoms $j$, deviates from initial bond
vector $\vec{r}^{\;0}_{ij}$,
\[
\vec{r}_{ij}\approx\left(1+\underline{\bm{\varepsilon}}\right){\vec{r}}_{ij}^{\;0}.\]
A SWNT can be viewed as a cylinder with radius $R$, its surface can
be perfectly embedded by six-member carbon rings \cite{oy97}. There
are three bond curves passing one carbon atoms at the surface of
SWNT, in the continuum limit, bond vector can be written
as~\cite{oy97}
\begin{eqnarray}
\vec{r}^{\;0}(M)&=&\vec{r}_{ij}^{\;0}=\left[1-a_{0}^{2}
\kappa^{2}(M)/6\right]\,a_{0}\,
\vec{t}(M)\nonumber\\
&+&\left[a_{0}\kappa(M)/2+a_{0}^{2}\kappa_{s}(M)/6\right]\,a_{0}\,\vec{N}(M)\\
&+&\left[\kappa(M)\tau(M)a_{0}^{2}/6\right]\,a_{0}\,\vec{b}(M),\nonumber
\end{eqnarray}
where $a_{0}=1.42\AA$ is carbon-carbon bond length without strains,
$M=1,2,3$ denote three $sp^{2}-$bonded curves from atom $i$ to atoms
$j$ on the surface of SWNT. Vectors $\vec{t},\;\vec{N}$ and
$\vec{b}$ are unit tangential, normal, and binormal vectors of the
bond curves from atom $i$ to $j$, $\kappa$, $\tau$ and $s$ are the
curvature, torsion, and arc parameter of bond curve, respectively,
$\kappa_{s}\equiv d\kappa/ds$.~\cite{oy97} The vectors
$\vec{t}(M)=\cos\theta(M)\hat{e}_{x}+\sin\theta(M)\vec{e}_{y},$
$\vec{b}(M)=\sin\theta(M)\hat{e}_{x}-\cos\theta(M)\hat{e}_{y},$
where $\hat{e}_{x}$ and $\hat{e}_{y}$ are the unit axial and
circumferential vectors at the $i$-atom's site on the SWNT surface,
$\theta(M)$ is the rotating angle from $\hat{e}_{x}$ to tangent
vector $\vec{t}$, which is related to the chiral angle $\theta_c$.~\cite{tu02} After deforming, bond length
$r_{ij}=|\vec{r}_{ij}|,$ and bond angle between bond vectors
$\vec{r}_{ij}$ and $\vec{r}_{ik}$ are
$\cos\theta_{ijk}=\vec{u}_{ij}\cdot\vec{u}_{ik},$ with unit vector
$\vec{u}_{ij}\equiv\vec{r}_{ij}/r_{ij}.$ 
Based on these relations, the $2$D continuum limit of elastic energy
per unit area of SWNT in Eq.~\ref{rebo}, which avoids introducing
ill-defined thickness of SWNTs,  can be written as,
\begin{equation}\label{elastic_energy}
\mathcal{E}_{elsticity}=\frac{1}{2}\sum_{ij}c_{ij}\varepsilon_{i}\varepsilon_{j}+\sum_{i\leq
j\leq k}c_{ijk}\varepsilon_{i}\varepsilon_{j}\varepsilon_{k},
\end{equation}
where $c_{ij}$ and $c_{ijk}$ are in-plane elastic
constants, $i,j,k=1,2,6$, they have analytical
expressions, see Appendix. Among them, $c_{16}$, harmonic
elastic constant for coupling between axial strain and
torsional twist is proportional to
$\left(a_0/R\right)^2\,\sin(6\theta_c)$, which clearly
shows a-SIT response is curvature and chirality effect,
only occurs in chiral SWNT. Obviously linear a-SIT
response is distinct at $\theta_c=\pi/12$ and significant
for SWNTs with small diameters, which are in accord with
previous theoretical and simulation works. For tubes with
large diameters and small strains, anharmonic elastic
energy can be ignored along with $c_{16}$ and $c_{26}$
terms, then the isotropic thin shell model for SWNTs is
recovered, and the calculated in-plane Young's modulus and
Possion's ratio are similar to the results in
Ref.~\onlinecite{tu02}.

To study the asymmetric a-SIT, we consider a chiral SWNT
with one fixed end, while the other end atoms are allowed
to relax both radially and tangentially during
deformation. The axial displacement is fixed for each
simulation step, ensuring that only axial stress occurs,
which is the basic assumption in simulations for a-SIT in
SWNTs~\cite{liang06,geng06,upmanyu08}.

The free energy per unit area of SWNT under axial stress is,
\begin{equation}
\mathcal{F}=\mathcal{E}_{elsticity}-\sigma_{1}\varepsilon_{1}.
\end{equation}

\begin{figure}[ht]
\scalebox{1.2}{\includegraphics{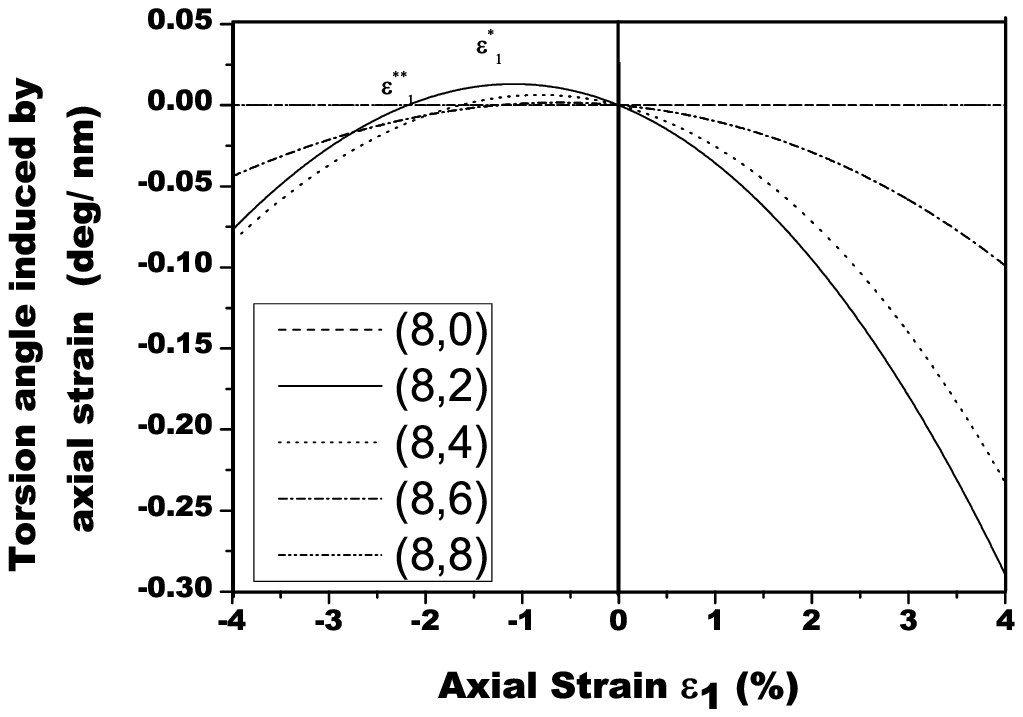}} \caption{\label{fig1}
Torsion angle-axial strain relations for a series of $(8,m)$ SWNTs,
which shows chirality dependence of a-SIT response. Only chiral
SWNTs have a-SIT response, as shown.}
\end{figure}

Assumption of equilibrium state leads to the following nonlinear equations
\begin{equation}
\frac{\partial\mathcal{F}}{\partial\varepsilon_{i}}=0,
\end{equation}
They give the relation between torsion angle per nm (in
unit of degree)
$\phi=-\left(180/\pi\right)\times\varepsilon_{6}/\left(R/1\mathrm{nm}\right)$
and axial strain $\varepsilon_{1}$, which is an asymmetric
response. There are two critical compressing strains
$\varepsilon_1^{*}$ and $\varepsilon_1^{**}$, as shown in
Fig.\ref{fig1}. For axial compression, at
$\varepsilon_{1}^{*}$, torsion angle reaches its extreme,
then SWNT begins to untwist, after totally untwisting at
critical strain $\varepsilon_{1}^{**}$, the tube twists
again to the opposite direction, i.e., to the direction as
the same as that for tension case.

\begin{figure}[ht]
\scalebox{1.2}{\includegraphics{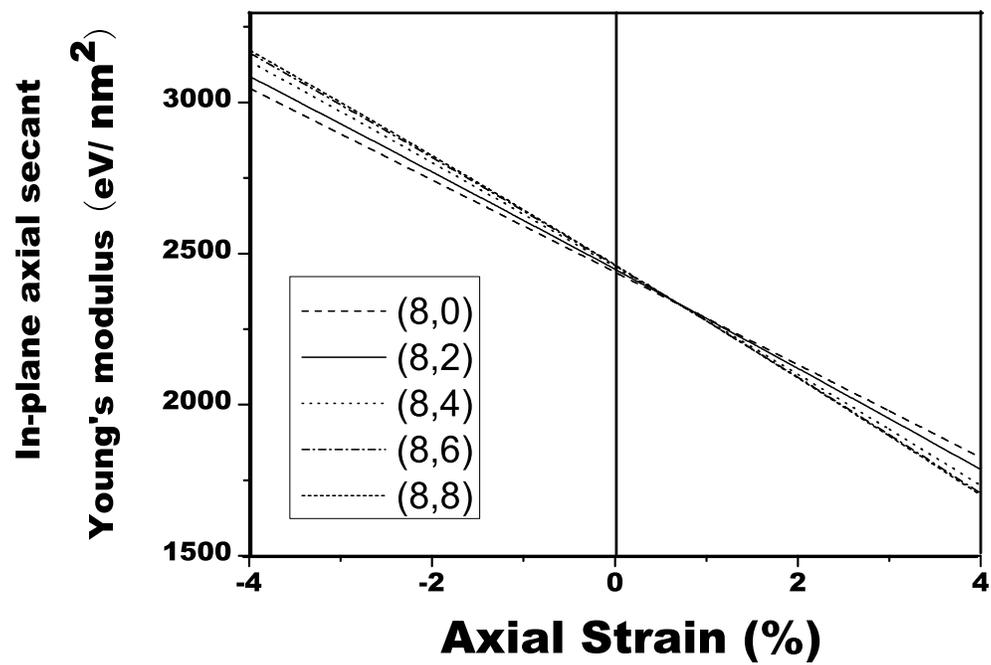}} \caption{\label{fig2} ' Relation between in-plane axial secant Young's modulus and axial strain, for a
series of $(8,m)$ SWNTs.}
\end{figure}

Another interesting result is nonlinear axial stress-strain
relation, the axial secant Young's modulus $Y_s \equiv d
\sigma/\,d\varepsilon_1$ of SWNT is a strict monotonically decreasing
function, $Y_{s}=Y_{s\,0}-t\,\varepsilon_{1}$, as shown in Fig. \ref{fig2}, thus SWNTs show
strain softening under tension, while strain hardening under compression.
This phenomenon was also found in recent molecular dynamic
simulations~\cite{geng06}.

We find asymmetric a-SIT and nonlinear axial stress-strain
of SWNT are tightly related to each other, the nature of
which is anharmonicity of atom-atom interaction for SWNTs,
such as REBO-II potential in Ref.~\onlinecite{liang06} and
present work. This anharmonicity leads to anharmonic bond
stretching energy in Eq.~\ref{rebo} and cubic terms in
Eq.~\ref{elastic_energy}, elastic energy. 

To illustrate it, we start from a simplified linear elastic energy
per unit area of SWNT,
\begin{equation}
\tilde{\mathcal{F}}=\frac{1}{2}c_{11}\varepsilon_{1}^{2}+
c_{16}\varepsilon_{1}\varepsilon_{6}+\frac{1}{2}c_{66}\varepsilon_{6}^{2}
-\sigma_{1}\varepsilon_{1},
\end{equation}
After substituting nonlinear stress-strain relation
$\sigma_{1}=Y_{s\,0}\,\varepsilon_{1}-\left(t^{2}/2\right)\,\varepsilon_{1}^{2}$
to $\tilde{\mathcal{F}}$, using equilibrium condition
$\partial\tilde{\mathcal{F}}/\partial\varepsilon_{1}=0,$
the torsion angle, which is proportional to
$\varepsilon_6$, is a quadratic function of axial strain.
$\phi\left(\varepsilon_{1}\right)$ curve is a parabola
with its symmetric axial located at $\varepsilon_{1}<0.$
Thus, present analysis captures main features of asymmetry
a-SIT.

Eq.~\ref{elastic_energy} without cubic terms gives the linear a-SIT
response's coefficient
\begin{equation}
\frac{d\phi}{d\varepsilon_{1}}|_{_{{\tiny
\varepsilon_{1}=0}}}=\frac{c_{12}\,c_{26}-c_{22}\,c_{16}}{c_{22}\,c_{66}}
\end{equation}
with leading term $\sim R^{-3}$ characterizing linear
a-SIT response, which is in good agreement with Gartstein
{\it et al}'s theoretical results. Therefore present
analysis captures the main characters of a-SIT response.

In our continuum elastic theory, we only get symmetric a-SIT without
anharmonic terms, however Ref.~\onlinecite{upmanyu08} gave asymmetric a-SIT by finite element simulation based on harmonic elasticity, although much small ($\sim$ $1/1000$ of those of molecular simulations). It may be due to the elastic energy we used is the
continuum limit of lattice energy, which may lose some
subtle microscopic information. Our continuum elastic theory
has some advantages, compared to previous simulations, for it is suitable to study SWNTs with actural sizes, and all the elastic constants
in the theory are obtained analytically. Obviously, our method can be
extended to calculate other anharmonic properties of SWNTs with
arbitrary radius and chiralities.

In summary, we emphasize the anharmonicity of inter-atoms
interactions and curvature-chirality induced anisotropic
elasticity are both important in a-SIT response, and
explain the asymmetry a-SIT and nonlinear stress-strain
relation all together. We find the unusual asymmetric
a-SIT effects is the consequence of the
curvature-chirality effect and anharmonic elasticity. We
give the analytical expressions of anharmonic elastic
energy, as well as curvature-chirality induced anisotropic
elasticity based on REBO-II. The calculated results are in
reasonable agreement with recent molecular dynamic
simulations. Our method can be used to analytically
calculate anharmonic properties of SWNTs with arbitrary
radius and chirality.

We are grateful for helpful discussion with Dr. H. Liang, Prof. Y.
Wang and Prof. J. Yan. We appreciate Dr. Hangtao Lu for his
carefully reading of the manuscripts.

\newpage
\newpage
\newpage
\section*{Appendix}
Elastic constants $c_{11},c_{12},\ldots,c_{666}$ presented
in Eq.~\ref{elastic_energy} can be described by,
\[\mathbf{cc}=\mathbf{M}\mathbf{b},\]
where, $\mathbf{cc}$ is a column vector with the
components $cc_1$ to $cc_{16}$ being the sixteen elastic
constants of SWCNT, i.e., $c_{11}$ to $c_{666,}$
respectively, $\mathbf{M}$ is a $16\times6$ matrix,
$\mathbf{b}$ is a column vector with components,
\[
b_1=\left(\frac{\partial^{2}V}{\partial
r_{ij}^{2}}\right)_{0}\cdot\frac{a_{0}^{2}}{\Omega_{0}},\;
b_2=\left(\frac{\partial^{2}V}{\partial\left(\cos\theta_{ijk}
\right)^{2}}\right)_{0}\cdot\frac{1}{\Omega_{0}},\;\]

\[
b_3=\left(\frac{\partial^{2}V}{\partial
r_{ij}\partial\cos\theta_{ijk}}\right)_{0}\cdot\frac{a_{0}}{\Omega_{0}},
b_4=\left(\frac{\partial^{2}V}{\partial\cos\theta_{ijk}\partial\cos
\theta_{ijl}}\right)_{0}\cdot\frac{1}{\Omega_{0}},\;
\]

\[ b_5=\left(\frac{\partial^{3}V}{\partial
r_{ij}^{3}}\right)_{0}\cdot\frac{a_{0}^{3}}{\Omega_{0}}.\]

Here, $a_0=1.42\AA$ is carbon-carbon bond length without
strains, and $\Omega_0=2.6\AA^2$ is the area occupied by
one carbon atom at the surface of SWCNTs.

All $34$ $ $non-zero elements of matrix $\mathbf{M}$ are
analytically written as ,
\begin{eqnarray*}
M_{1,1} & = &
\frac{9}{16}+\left(\frac{-45}{1024}\right)\alpha^{2},\\
M_{2,1}&=&\frac{3}{16}+\left(\frac{-43}{1024}\right)\alpha^{2}+
\left(\frac{-11}{256}\right)\alpha^{2}\cos(6\theta),\\
M_{3,1} & = &
\frac{9}{16}+\left(\frac{-205}{1024}\right)\alpha^{2}+
\left(\frac{11}{128}\right)\alpha^{2}\cos(6\theta),\\
M_{4,1}&=&\left(\frac{11}{512}\right)\alpha^{2}\sin(6\theta),\\
M_{5,1} & = & \left(\frac{-33}{512}\right)\alpha^{2}\sin(6\theta),\\
M_{6,1}&=&\frac{3}{16}+\left(\frac{-43}{1024}\right)\alpha^{2}+
\left(\frac{-11}{256}\right)\alpha^{2}\cos(6\theta),\\
M_{1,2} & = &
\frac{27}{16}+\left(\frac{-27}{64}\right)\alpha^{2}
+\left(\frac{-189}{512}\right)\alpha^{2}\cos(6\theta),\\
M_{2,2}&=&\frac{-27}{16}+\left(\frac{27}{128}\right)\alpha^{2}
+\left(\frac{27}{64}\right)\alpha^{2}\cos(6\theta),\\
M_{3,2} & = & \frac{27}{16}+\left(\frac{-243}{256}\right)\alpha^{2}\cos(6\theta),\\
M_{4,2}&=&\left(\frac{-405}{1024}\right)\alpha^{2}\sin(6\theta),\\
M_{5,2} & = & \left(\frac{459}{1024}\right)\alpha^{2}\sin(6\theta),\\
M_{6,2}&=&\frac{27}{16}+\left(\frac{-27}{256}\right)\alpha^{2}
+\left(\frac{27}{64}\right)\alpha^{2}\cos(6\theta),\\
M_{1,3} & = &
\frac{-9}{8}+\left(\frac{-285}{1024}\right)\alpha^{2}
+\left(\frac{-3}{64}\right)\alpha^{2}\cos(6\theta),\\
M_{2,3}&=&\frac{9}{8}+\left(\frac{-627}{1024}\right)\alpha^{2}
+\left(\frac{-3}{256}\right)\alpha^{2}\cos(6\theta),\\
M_{3,3} & = &
\frac{-9}{8}+\left(\frac{-189}{1024}\right)\alpha^{2}+
\left(\frac{9}{128}\right)\alpha^{2}\cos(6\theta),\\
M_{4,3}&=&\left(\frac{-9}{512}\right)\alpha^{2}\sin(6\theta),\\
M_{5,3} & = & \left(\frac{-21}{512}\right)\alpha^{2}\sin(6\theta),\\
M_{6,3}&=&\frac{-9}{8}+\left(\frac{165}{1024}\right)\alpha^{2}+
\left(\frac{-3}{256}\right)\alpha^{2}\cos(6\theta),\\
M_{1,4} & = &
\frac{-27}{32}+\left(\frac{27}{128}\right)\alpha^{2}
+\left(\frac{189}{1024}\right)\alpha^{2}\cos(6\theta),\\
M_{2,4}&=&\frac{27}{32}+\left(\frac{-27}{256}\right)\alpha^{2}+
\left(\frac{-27}{128}\right)\alpha^{2}\cos(6\theta),\\
M_{3,4} & = & \frac{-27}{32}+\left(\frac{243}{1024}\right)\alpha^{2}\cos(6\theta),\\
M_{4,4}&=&\left(\frac{-459}{2048}\right)\alpha^{2}\sin(6\theta),\\
M_{5,4} & = & \left(\frac{-459}{2048}\right)\alpha^{2}\sin(6\theta),\\
M_{6,4}&=&\frac{-27}{32}+\left(\frac{27}{512}\right)\alpha^{2}+
\left(\frac{-27}{128}\right)\alpha^{2}\cos(6\theta),\\
M_{7,5} & = & \frac{5}{64}+\left(\frac{1}{128}\right)\cos(6\theta),\\
M_{8,5}&=&\frac{3}{64}+\left(\frac{-3}{128}\right)\cos(6\theta),\\
M_{9,5} & = & \left(\frac{3}{128}\right)\sin(6\theta),\\
M_{10,5}&=&\frac{3}{64}+\left(\frac{3}{128}\right)\cos(6\theta),\\
M_{11,5} & = & \left(\frac{-3}{64}\right)\sin(6\theta),\\
M_{12,5}&=&\frac{3}{64}+\left(\frac{-3}{128}\right)\cos(6\theta),\\
M_{13,5} & = & \left(\frac{5}{64}\right)+\left(\frac{-1}{128}\right)\cos(6\theta),\\
M_{14,5}&=&\left(\frac{3}{128}\right)\sin(6\theta),\\
M_{15,5} & = &
\left(\frac{3}{64}\right)+\left(\frac{3}{128}\right)\cos(6\theta),\\
M_{16,5}&=&\left(\frac{-1}{128}\right)\sin(6\theta).
\end{eqnarray*}
Here, $\alpha\equiv a_0/R$.
\end{document}